\documentclass[aps,prc,floatfix,nofootinbib,showpacs,twocolumn,superscriptaddress]{revtex4-1}

\usepackage{latexsym}
\usepackage{amsmath}
\usepackage{amssymb}
\usepackage{amsfonts}
\usepackage{pzccal}

\usepackage{color}

\usepackage{supertabular}
\usepackage{placeins}
\usepackage{epsfig}
\usepackage{graphicx}

\definecolor{darkgreen}{rgb}{0,0.5,0}
\definecolor{purple}{rgb}{0.5,0,0.5}
\definecolor{blue}{rgb}{0.0,0.0,0.50}
\definecolor{scarlet}{rgb}{1.0,0.2,0}

\begin{document}


\title{Dissecting nucleon transition electromagnetic form factors}



\author{Jorge Segovia}
\email[]{jorge.segovia@tum.de}
\affiliation{Physik-Department, Technische Universit\"at M\"unchen,
James-Franck-Str.\,1, D-85748 Garching, Germany}

\author{Craig D. Roberts}
\email[]{cdroberts@anl.gov}
\affiliation{Physics Division, Argonne National Laboratory, Argonne, Illinois
60439, USA}

\date{14 July 2016}

\begin{abstract}
In Poincar\'e-covariant continuum treatments of the three valence-quark bound-state problem, the force behind dynamical chiral symmetry breaking also generates nonpointlike, interacting diquark correlations in the nucleon and its resonances.  We detail the impact of these correlations on the electromagnetically-induced nucleon-$\Delta$ and nucleon-Roper transitions, providing a flavour-separation of the latter and associated predictions that can be tested at modern facilities.
\end{abstract}



\maketitle



\noindent\textbf{1.$\;$Introduction}.
Owing to asymptotic freedom, QCD is well defined in the absence of a Lagrangian current-quark mass; and the appearance of a nonzero quark running-mass in this (chiral) limit is a characteristic signal of dynamical chiral symmetry breaking (DCSB), which has numerous observable consequences \cite{Cloet:2013jya, Roberts:2015lja, Horn:2016rip}.  An important but lesser known corollary of DCSB is the appearance of strong quark-quark (diquark) correlations within baryons \cite{Cahill:1987qr, Cahill:1988dx, Burden:1988dt, Maris:2002yu, Segovia:2015ufa, Santopinto:2016zkl, Eichmann:2016yit}.  These correlations are nonpointlike and fully interacting, and each of a baryon's three dressed-quarks is involved in every type of correlation to the fullest extent allowed by its quantum numbers and those of the bound state.  One should therefore expect the spectrum obtained in the presence of such correlations to be as rich as that allowed by a three-constituent quark model; 
and this expectation is supported by existing calculations \cite{Chen:2012qr, Eichmann:2016yit}.  On the other hand, it has been argued that the presence of diquark correlations has a dramatic effect on the contribution from different quark flavours to, \emph{inter alia},  nucleon elastic electromagnetic form factors \cite{Cloet:2008re, Segovia:2015ufa} and parton distribution functions \cite{Roberts:2013mja}; and, where data exists, the associated predictions are confirmed \cite{Cates:2011pz, Qattan:2012zf, Parno:2014xzb}.

The same picture has also been employed in the description of nucleon-to-resonance transition form factors, where the final state is either the $\Delta$-baryon [$\Delta(1232)(3/2)^+$] or Roper resonance [$N(1440)(1/2)^+$] \cite{Eichmann:2011aa, Segovia:2013rca, Segovia:2014aza, Segovia:2015hra}.  This suggests the absence of environment sensitivity for DCSB in the nucleon, $\Delta$ and Roper, \emph{i.e}.\ DCSB in these systems is expressed in ways that can readily be predicted once its manifestation is understood in the pion, and this includes the generation of diquark correlations with the same character in each of these baryons \cite{Roberts:2016dnb}.  In order to aid the empirical validation of these ideas, herein we follow Refs.\,\cite{Segovia:2014aza, Segovia:2015hra} and present a comprehensive analysis of the transition form factors via their separation into contributions from different correlation sectors and subsequently, where appropriate, a flavour separation for each of these.  

\smallskip



\noindent\textbf{2.$\;$Baryon Structure and Electromagnetic Transition Currents}.
In relativistic quantum field theory, a baryon is described by a Faddeev amplitude, obtained from a Poincar\'e-covariant Faddeev equation that sums all possible exchanges and interactions that can take place between the three dressed-quarks that express its valence-quark content \cite{Segovia:2014aza, Segovia:2015hra}.  A dynamical prediction of Faddeev equation studies that employ realistic quark-quark interactions \cite{Binosi:2014aea} is the appearance of nonpointlike quark$+$quark (diquark) correlations within baryons, whose characteristics are determined by DCSB \cite{Segovia:2015ufa}.  
In order to understand the discussion herein, it is important to bear in mind that, using realistic Faddeev kernels, the nucleon ($N$) and Roper-resonance ($R$) contain both scalar and pseudovector diquarks, with the scalar diquark contributing 62\% of their normalisation \cite{Segovia:2015hra}, but the $\Delta$-baryon involves only pseudovector diquarks because it is impossible to form an isospin-$3/2$ baryon from a quark and isoscalar-scalar diquark \cite{Oettel:1998bk, Alkofer:2004yf}.


Electromagnetically induced $N\to \Delta$, $N\to R$ transitions proceed via the current introduced in Ref.\,\cite{Oettel:1999gc} and refined in Ref.\,\cite{Segovia:2014aza} (see Appendix~C therein).  In two separate ways, this current can be considered as a sum of three distinct terms, \emph{viz}.
\begin{description}
\item[T1 = diquark dissection] \emph{T1A} --
 sca\-lar diquark in both the initial- and final-state baryon,
\emph{T1B} -- pseudovector diquark in both the initial- and final-state baryon, and
\emph{T1C} -- a different diquark in the initial- and final-state baryon; and
\item[T2 = scatterer dissection]  \emph{T2A} -- photon strikes a bystander dressed-quark,
\emph{T2B} -- photon interacts with a diquark, elastically or causing a transition scalar\,$\leftrightarrow$\,pseudovector, and
\emph{T2C} -- photon strikes a dressed-quark in-flight, as one diquark breaks up and another is formed, or appears in one of the two associated ``seagull'' terms.
\end{description}
The anatomy of a given transition process is revealed by combining the information provided by the T1 and T2 dissections.

\smallskip

\noindent\textbf{3.$\;$\mbox{\boldmath $N\to \Delta$}}.
%
%
%
In the isospin symmetric limit, there is no difference between $\gamma + p \to \Delta^+$ and $\gamma + n \to \Delta ^0$.  Hence, a flavour separation of the associated form factors is impossible.  We focus, therefore, on $\gamma+p \to \Delta^+$, which has been measured on a domain of momentum transfers that extends to $Q^2 \approx 6\,$GeV$^2$ \cite{Bartel:1968tw, Stein:1975yy, Sparveris:2004jn, Stave:2008aa, Aznauryan:2009mx}.  This transition is described by three Poincar\'e-invariant form factors \cite{Jones:1972ky}: magnetic-dipole, $G_{M}^{\ast}$; electric quadrupole, $G_{E}^{\ast}$; and Coulomb (longitudinal) quadrupole, $G_{C}^{\ast}$.   Concerning $G_{M,C}^{\ast}$, it has been established that the baryons' dressed-quark cores are revealed to probes with $x_\Delta = Q^2/m_\Delta^2 \gtrsim (1/2)$ \cite{Segovia:2014aza, Roberts:2015dea}.  However, owing to the small magnitude of $G_{E}^{\ast}$, meson-baryon final-state interactions (MB\,FSIs or ``meson cloud'' effects) obscure the core on a larger domain, which likely extends to $x_\Delta \gtrsim 4$ \cite{Segovia:2014aza}.

\begin{figure}[t]
\centerline{%
\includegraphics[clip,width=0.39\textwidth]{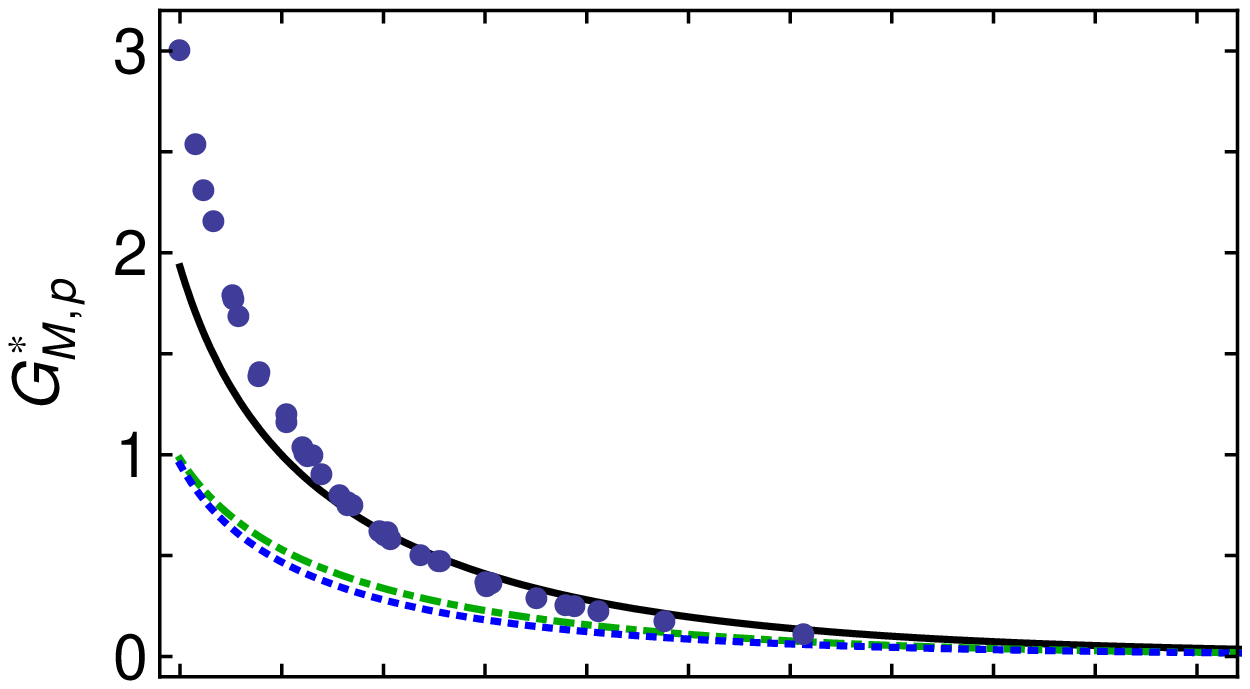}}
\vspace*{-11.1ex}

\centerline{%
\includegraphics[clip,width=0.39\textwidth]{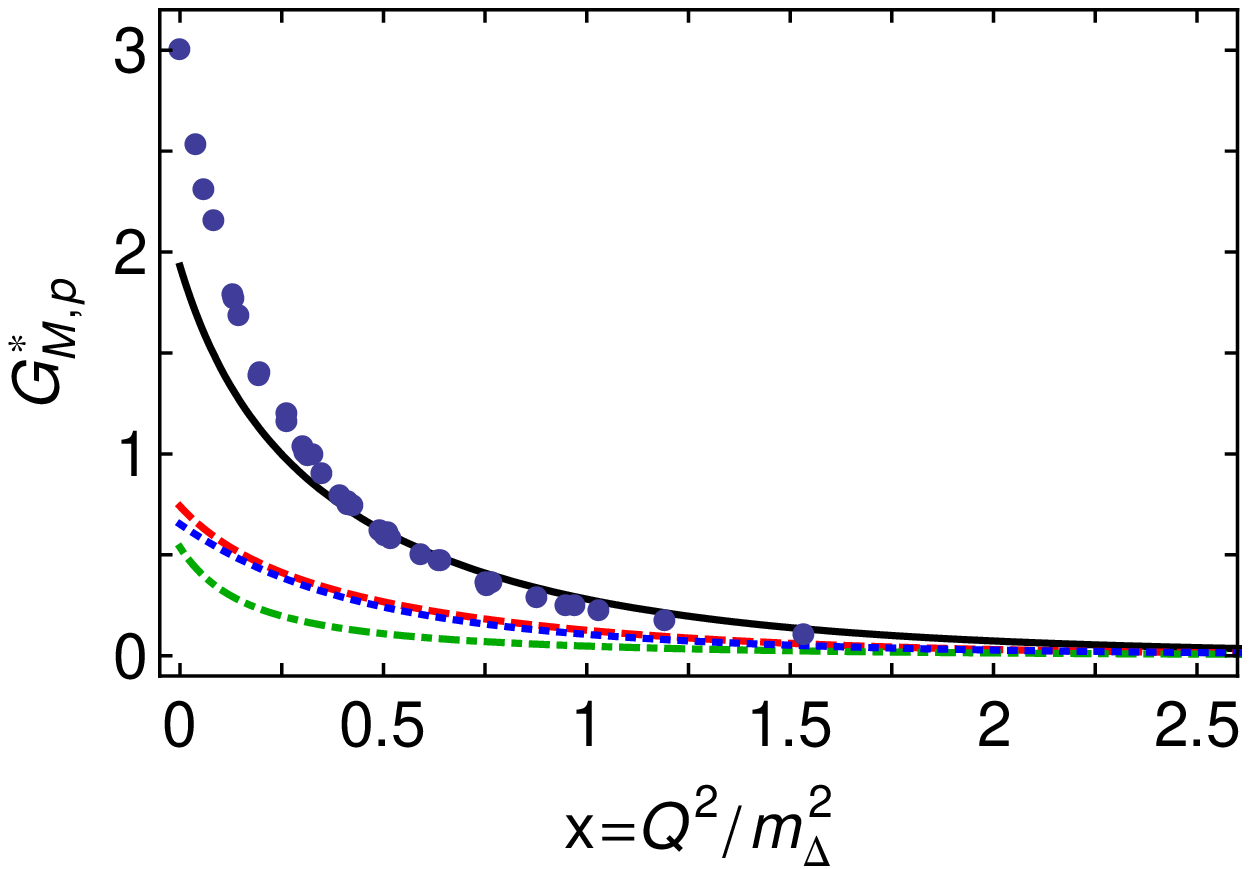}}
\caption{\label{figGM}
$\gamma\,p\to\Delta^+$ magnetic dipole form factor, $G_M^\ast$, computed as described in Ref.\,\cite{Segovia:2014aza} (solid black curve).  The mismatch at small-$x$ between data \cite{Bartel:1968tw, Stein:1975yy, Sparveris:2004jn, Stave:2008aa, Aznauryan:2009mx} and prediction is explained by meson-cloud effects \cite{Segovia:2014aza, Sato:2000jf, JuliaDiaz:2006xt, Kamano:2013iva}.
\emph{Upper panel} -- diquark dissection: \emph{T1B} (dot-dashed green), pseudovector diquark in both $p$, $\Delta^+$; \emph{T1C} (dotted blue), scalar diquark in $p$, pseudovector diquark in $\Delta^+$.
\emph{Lower panel} -- scatterer dissection: \emph{T2A} (red dashed), photon strikes an uncorrelated dressed quark; \emph{T2B} (dot-dashed green), photon strikes a diquark; and \emph{T2C} (dotted blue), diquark breakup contributions, including photon striking exchanged dressed-quark.
}
\end{figure}

The anatomy of $G_{M}^{\ast}$ is revealed in Fig.\,\ref{figGM}.  The upper panel shows that \emph{T1B} and \emph{T1C} diagrams contribute equally, but, since the scalar diquark, $[ud]$, is a larger part of the nucleon's Faddeev amplitude, this means that contributions with a pseudovector diquark, $\{qq\}$, in both $p$ and $\Delta^+$ contribute more strongly to the transition.
The lower panel shows that the dominant contributions are those in which the photon strikes a dressed-quark.
Hence, the magnetic component of the transition proceeds predominantly via spin-flip of an uncorrelated quark, \emph{T2C} for $[ud]$ in the proton and \emph{T2A} for $\{qq\}$, with slightly greater transition strength in the latter configuration.

\begin{figure}[t]
\centerline{%
\includegraphics[clip,width=0.4\textwidth]{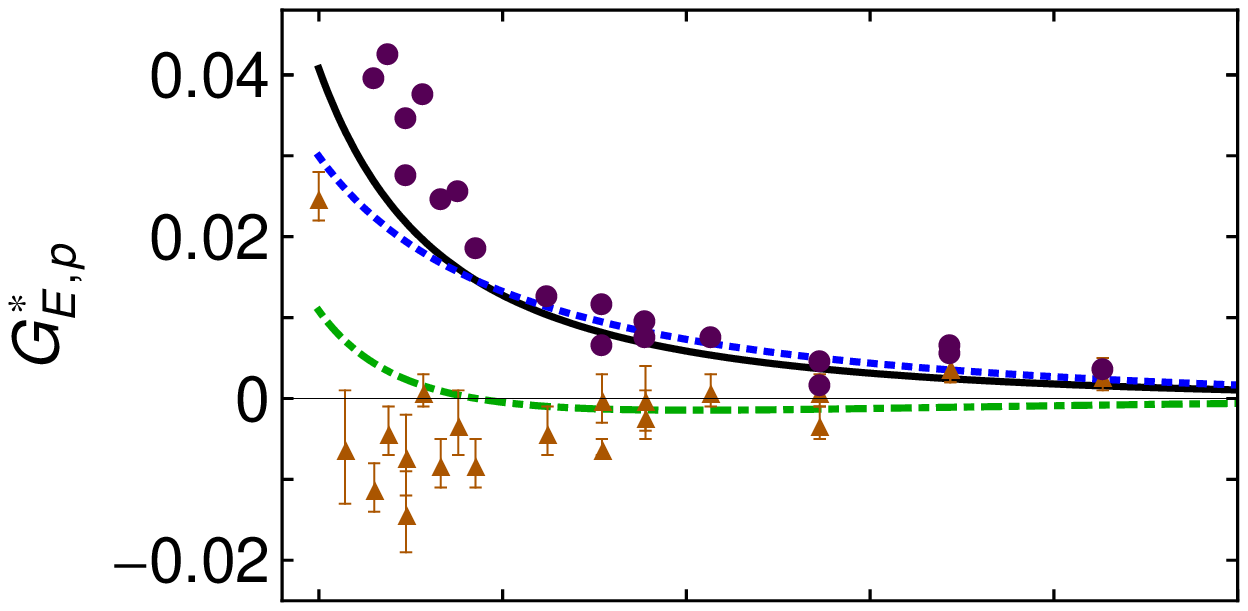}}
\vspace*{-11.3ex}

\centerline{%
\includegraphics[clip,width=0.4\textwidth]{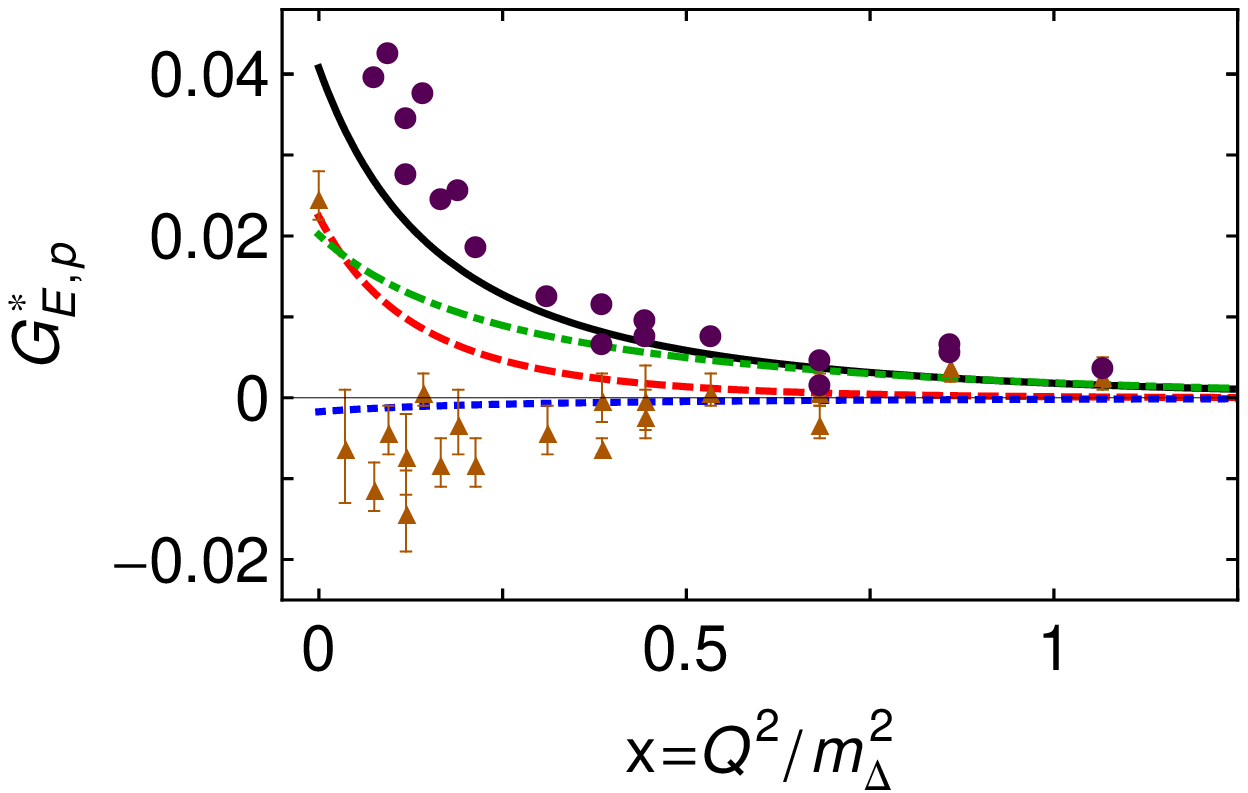}}
\caption{\label{figGE}
$\gamma\,p\to\Delta^+$ electric quadrupole form factor, $G_E^\ast$, computed as described in Ref.\,\cite{Segovia:2014aza} (solid black curve).  The ``data'' are drawn from the dynamical coupled channels analysis in Ref.\,\cite{JuliaDiaz:2006xt}: circles, complete result; and triangles, an estimate of $G_E^\ast$ as it would appear in the absence of MB\,FSIs.
The legend is otherwise as it appears in Fig.\,\ref{figGM}.
%
%
}
\end{figure}

The electric quadrupole transition form factor, $G_{E}^{\ast}$, is dissected in Fig.\,\ref{figGE}.  The upper panel shows that this component of the transition is dominated by diagrams involving a scalar diquark in the proton and a pseudovector diquark in the $\Delta^+$ (\emph{T1C}); and the lower panel indicates that photon-diquark interactions control the transition away from $x_\Delta \simeq 0$.  It follows that, within the dressed-quark core, the electric quadrupole transition proceeds primarily by a photon transforming the $0^+$-diquark into a $1^+$-diquark ($\delta J=1$) with the overlap of what may be said to be quark-diquark components in the rest-frame Faddeev wave functions of the proton and $\Delta^+$ that differ by one unit of angular momentum.

Whilst not apparent in Fig.\,\ref{figGE}, $G_{E}^{\ast}$ possesses a zero at $x_\Delta\approx 3$ (see Fig.\,11 in Ref.\,\cite{Segovia:2014aza}).  Its origin lies in the existence of a zero at $x_\Delta\approx 3$ in the (\emph{T1C}) $[ud] \to \{ud\}$ diquark transition contribution.  However, its position is also influenced by the size of the dressed-quark anomalous magnetic moment ($\kappa_Q$) \cite{Singh:1985sg, Bicudo:1998qb, Kochelev:1996pv, Chang:2010hb}, shifting to larger values of $x_\Delta$ as this moment increases.  Our analysis reveals the cause, \emph{viz}.\ greater values of $\kappa_Q$ increase the domain of positive support for the \emph{T2A} contribution, and this pushes the zero to larger values of $x_\Delta$.  The impact of $\kappa_Q$ is also a signal that meson-cloud effects are important to the behaviour of $G_{E}^{\ast}$ because they act, \emph{inter alia}, to increase the magnitude of $\kappa_Q$ \cite{Horikawa:2005dh, Cloet:2012cy, Cloet:2014rja}.

It is worth recalling here that helicity conservation at large momentum scales requires $R_{\rm EM}(x_\Delta) = -G_E^{\ast}/G_M^{\ast} \to 1$ as $x_\Delta\to \infty$ \cite{Carlson:1985mm, Segovia:2013rca}, so $G_{E}^{\ast}$ must exhibit a zero once the dressed-quark core becomes the dominant contribution.  Apparently, however, this is not the case on the measured domain, within which many factors compete in producing $G_E^{\ast}$.  The analysis in Ref.\,\cite{Segovia:2014aza} suggests that the zero occurs on $x_\Delta \gtrsim 6$.

Figure\,\ref{figGE} also compares our calculation with the dynamical coupled-channels (DCC) analysis of $G_{E}^{\ast}$ in Ref.\,\cite{JuliaDiaz:2006xt}.  In contrast to $G_M^\ast$ in Fig.\,\ref{figGM}, here the DCC-inferred bare contribution (triangles) is zero, within errors, and hence differs markedly from our prediction for the dressed-quark core contribution.  Instead, the complete DCC result (circles) is aligned with our curve.  Given that Poincar\'e-covariance entails the presence of rest-frame quark-diquark orbital angular momentum in bound-state Faddeev amplitudes, we judge that our result is the better estimate of the dressed-quark core contribution in this case, with the analysis in Ref.\,\cite{JuliaDiaz:2006xt} failing here owing to the very small size of $G_{E}^{\ast}$ on the measured domain.

\begin{figure}[t]
\centerline{%
\includegraphics[clip,width=0.375\textwidth]{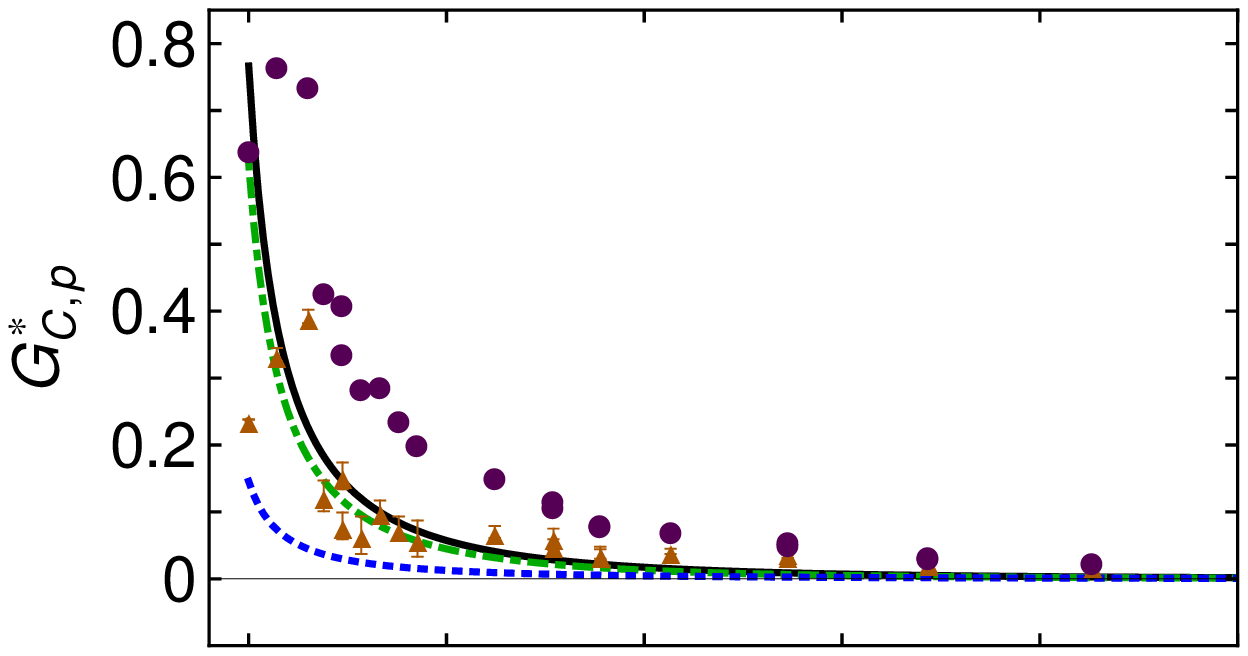}}
\vspace*{-10.7ex}

\centerline{%
\includegraphics[clip,width=0.375\textwidth]{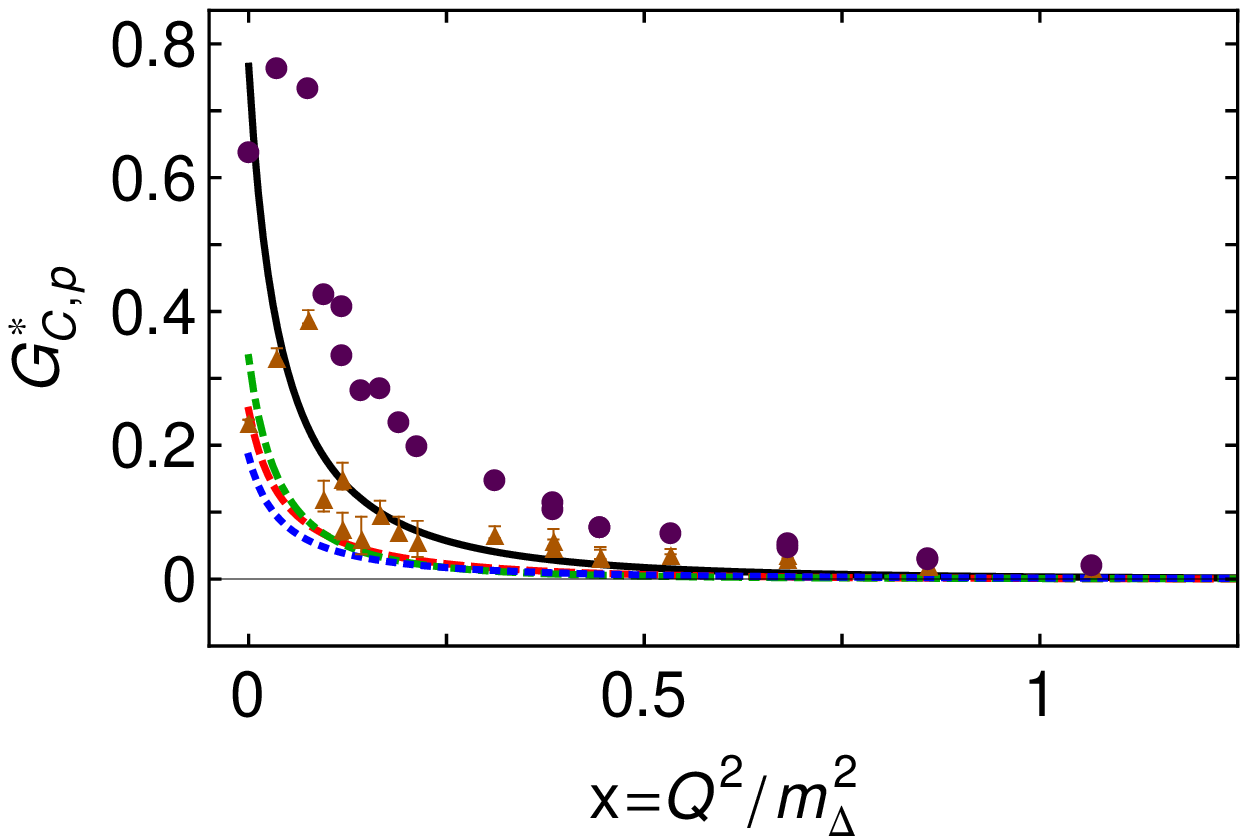}}
\caption{\label{figGC}
$\gamma\,p\to\Delta^+$ Coulomb quadrupole form factor, $G_C^\ast$, computed as described in Ref.\,\cite{Segovia:2014aza} (solid black curve).
%
%
The legend is as it appears in Fig.\,\ref{figGE}, with the results from Ref.\,\cite{JuliaDiaz:2006xt} multiplied by ``$-1$'' in order to match our conventions for this form factor.
%
%
}
\end{figure}

The anatomy of $G_{C}^{\ast}$ is revealed in Fig.\,\ref{figGC}.  The upper panel indicates that \emph{T1B} contributions dominate, \emph{i.e}.\ transition processes with a pseudovector diquark in both the proton and $\Delta^+$.  The lower panel shows that photon-diquark interactions control the transition at small $x_\Delta$, but photon-quark scattering dominates on $x_\Delta \gtrsim 1/8$.  Regarding the dressed-quark core,  these observations suggest that on $x_\Delta \simeq 0$ this $C2$ transition is influenced a little by the strength of the pseudovector diquark's own quadrupole moment, but otherwise it is a measure of an overlap between what may be called $S$- and $D$-wave quark-diquark angular momentum components in the rest-frame proton and $\Delta^+$ Faddeev wave functions.

It is worth remarking that $G_{C}^{\ast}$ is an order-of-magnitude larger than $G_{E}^{\ast}$; and in this case, as with $G_{M}^{\ast}$, the DCC-inferred bare contribution to $G_{C}^{\ast}$, computed in Ref.\,\cite{JuliaDiaz:2006xt} and depicted in Fig.\,\ref{figGC}, agrees well with our prediction for the dressed-quark core component.

\begin{figure}[t]
\centerline{%
\includegraphics[clip,width=0.40\textwidth]{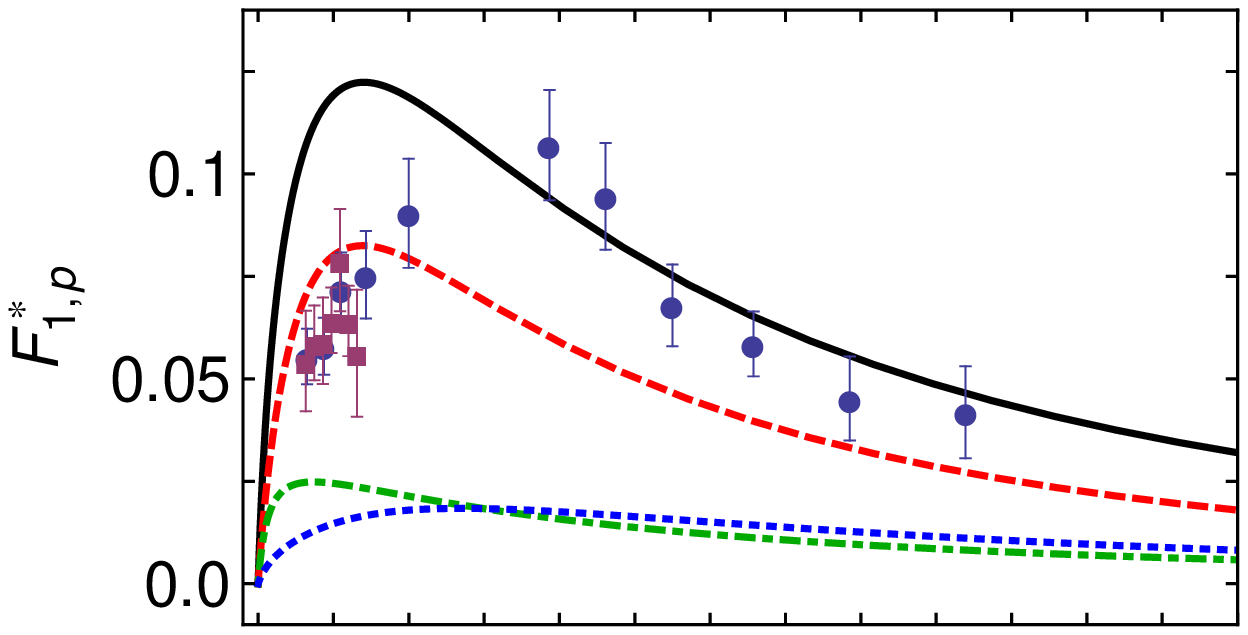}}
\vspace*{-11.3ex}

\centerline{%
\includegraphics[clip,width=0.40\textwidth]{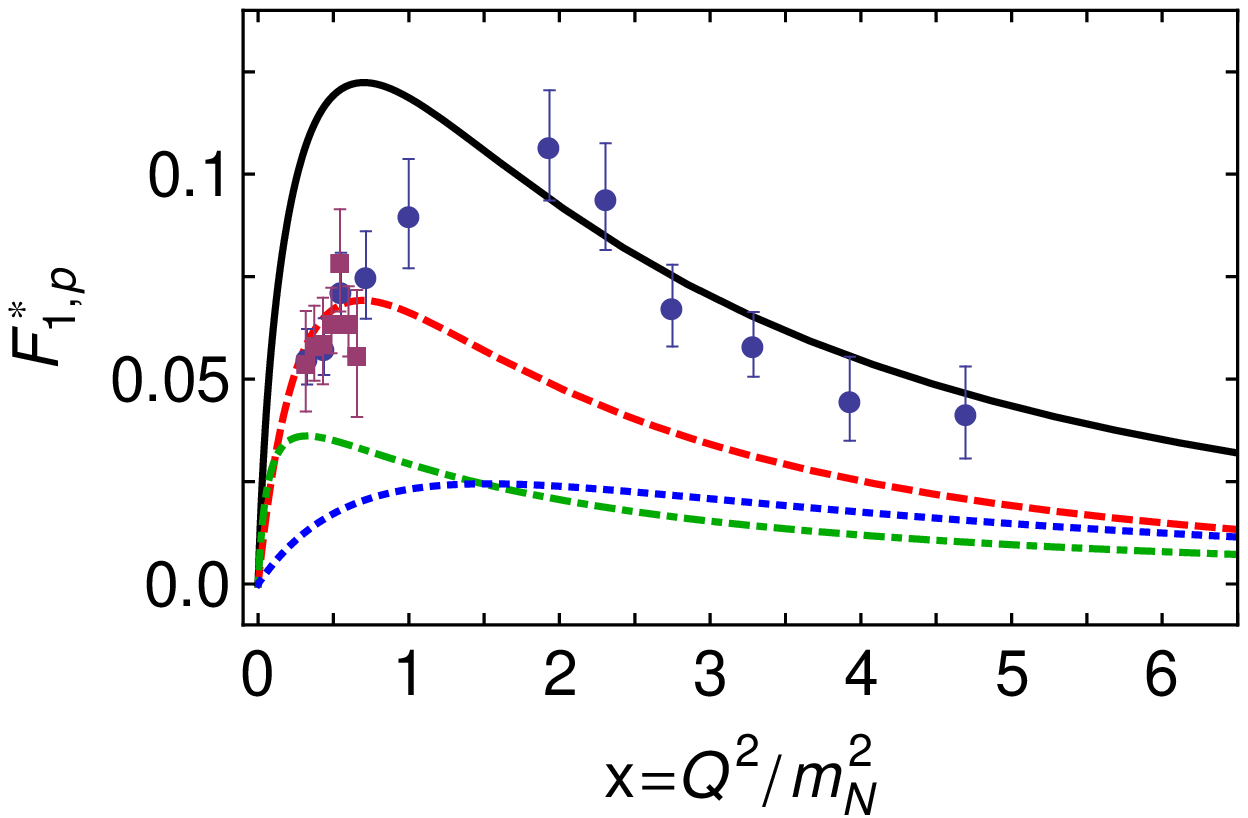}}
\caption{\label{figF1}
$\gamma\,p\to R^+$ Dirac transition form factor, $F_{1,p}^{\ast}$ as a function of $x=Q^2/m_N^2$, computed as described in Ref.\,\cite{Segovia:2015hra} (solid black curve).
Data: circles (blue) \cite{Aznauryan:2009mx} and squares (purple) \cite{Mokeev:2012vsa}.
\emph{Upper panel} -- diquark breakdown: \emph{T1A} (dashed red), scalar diquark in both $p$, $R^+$; \emph{T1B} (dot-dashed green), pseudovector diquark in both $p$, $R^+$; \emph{T1C} (dotted blue), scalar diquark in $p$, pseudovector diquark in $R^+$, and vice versa.
\emph{Lower panel} -- scatterer breakdown: \emph{T2A} (red dashed), photon strikes an uncorrelated dressed quark; \emph{T2B} (dot-dashed green), photon strikes a diquark; and \emph{T2C} (dotted blue), diquark breakup contributions, including photon striking exchanged dressed-quark.
}
\end{figure}

\smallskip

\noindent\textbf{4.$\;$\mbox{\boldmath $N\to R$}}.
The nature of the Roper resonance has long been controversial, see \emph{e.g}.\ Refs.\,\cite{Cardarelli:1996vn, Glozman:1997ag, Capstick:1999qq, Tiator:2003uu, JuliaDiaz:2006av, Aznauryan:2007ja, Nagata:2008hi, Edwards:2011jj, Lin:2011da, Wilson:2011aa, Aznauryan:2012ec, Engel:2013ig, Alexandrou:2013fsu, Leinweber:2015kyz}, but
a new case has recently been made \cite{Suzuki:2009nj, Segovia:2015hra, Roberts:2016dnb, Aznauryan:2016wwm} in support of the view that the observed Roper resonance is at heart the nucleon's first radial excitation, consisting of a well-defined dressed-quark core augmented by a meson cloud that reduces its (Breit-Wigner) mass by approximately 20\%.  As part of this explanation, a meson-cloud obscures the dressed-quark core from long-wavelength probes, but that core is revealed to probes with $x_N = Q^2/m_N^2 \gtrsim 3$.  Here we dissect the associated nucleon-Roper transition form factors and describe their flavour separation.  Since experiments have already yielded precise information on proton-Roper transition form factors \cite{Dugger:2009pn, Aznauryan:2009mx, Aznauryan:2011qj, Mokeev:2012vsa, Mokeev:2015lda, Burkert:2016dxc}, these predictions could be validated following electroproduction experiments on (bound-)\,neutron targets.

The anatomy of the $\gamma\,p\to R^+$ Dirac transition form factor is revealed in Fig.\,\ref{figF1}.  Plainly, this component of the transition proceeds primarily through a photon striking a bystander dressed quark that is partnered by $[ud]$, with lesser but non-negligible contributions from all other processes.  In exhibiting these features, $F_{1,p}^{\ast}$ shows marked qualitative similarities to the proton's elastic Dirac form factor (\emph{cf}.\, Fig.\,3 in Ref.\,\cite{Cloet:2008re}).

\begin{figure}[t]
\centerline{%
\includegraphics[clip,width=0.40\textwidth]{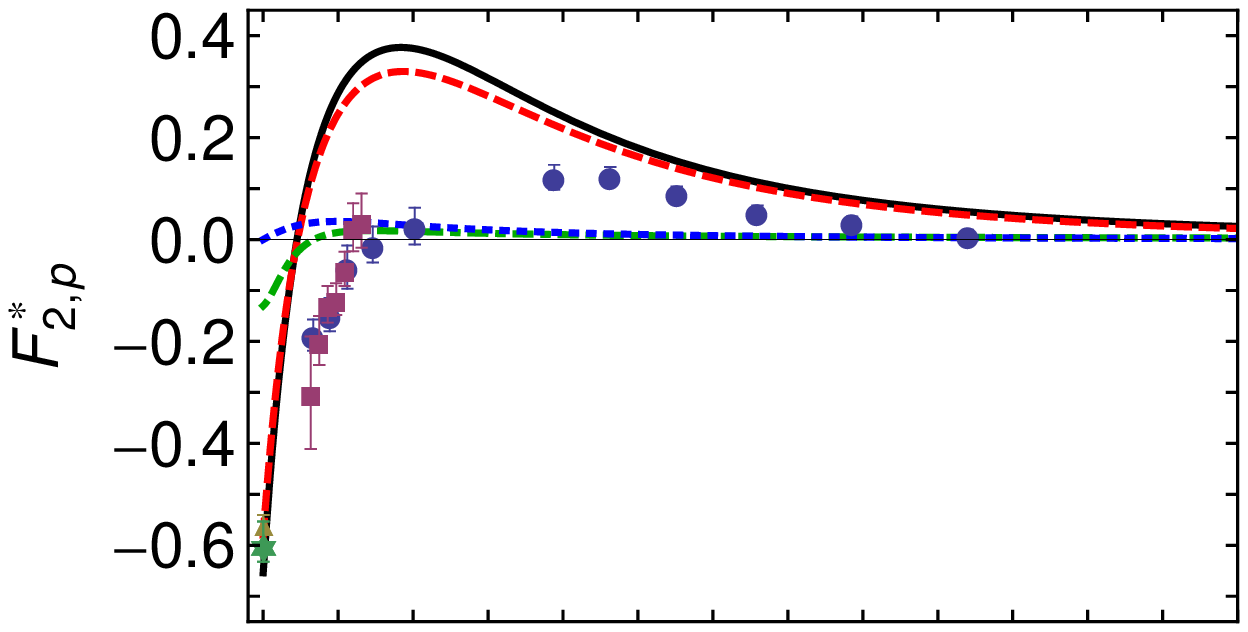}}
\vspace*{-11.3ex}

\centerline{%
\includegraphics[clip,width=0.40\textwidth]{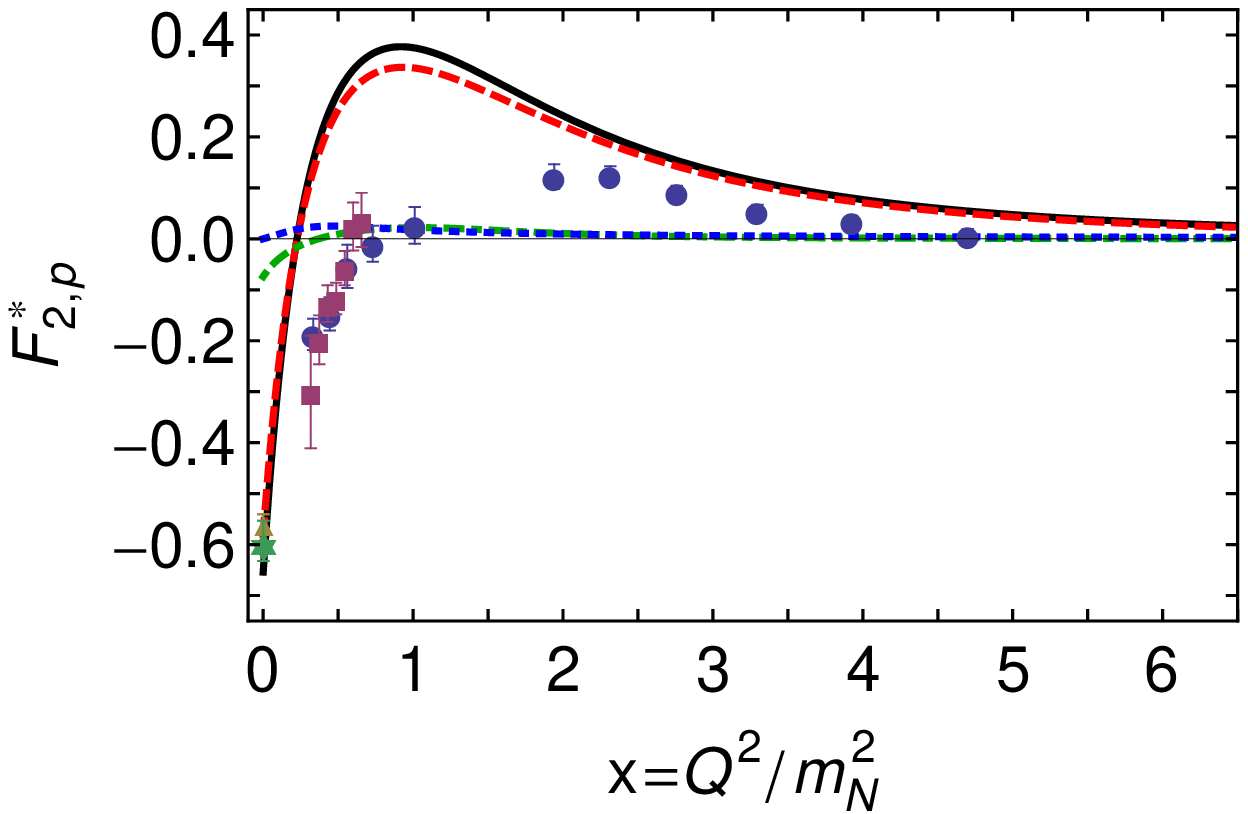}}
\caption{\label{figF2}
$\gamma\,p\to R^+$ Pauli transition form factor, $F_{2,p}^{\ast}$ as a function of $x=Q^2/m_N^2$, computed as described in Ref.\,\cite{Segovia:2015hra} (solid black curve).
Data:
circles (blue) \cite{Aznauryan:2009mx};
squares (purple) \cite{Mokeev:2012vsa};
triangle (gold) \cite{Dugger:2009pn};
and star (green) \cite{Agashe:2014kda}.
\emph{Upper panel} -- diquark breakdown; \emph{lower panel} -- scatterer breakdown; and legend as in Fig.\,\ref{figF1}.
%
%
}
\end{figure}

The $\gamma\,p\to R^+$ Pauli transition form factor is dis\-sected in Fig.\,\ref{figF2}. In this case, a single contribution is overwhelmingly important, \emph{viz}.\ photon strikes a bystander dressed-quark in association with $[ud]$ in the proton and $R^+$.  No other diagram makes a significant contribution.  A comparison with Fig.\,4 in Ref.\,\cite{Cloet:2008re} reveals that the same may be said for the dressed-quark core component of the proton's elastic Pauli form factor.

In hindsight, given that the diquark content of the proton and $R^+$ are almost identical, with the $\psi_0 \sim u[ud]$ component contributing roughly 60\% of the charge of both systems, the qualitative similarity between the proton elastic and proton-Roper transition form factors is not surprising.  This observation immediately raises the issue of whether and how that similarity is transmitted into the flavour separated form factors.

If one supposes that $s$-quark contributions to the nucleon-Roper transitions are negligible, as is the case for nucleon elastic form factors, and assumes isospin symmetry, then a flavour separation of the transition form factors is accomplished by combining results for the $\gamma\,p\to\, R^+$ and $\gamma\,n\to R^0$ transitions:
\begin{equation}
\label{FlavourSep}
F_{1(2),u}^{\ast} = 2 F_{1(2)}^{\ast,p} + F_{1(2)}^{\ast,n}, \;
F_{1(2),d}^{\ast} = 2 F_{1(2)}^{\ast,n} + F_{1(2)}^{\ast,p},
\end{equation}
where $p$ and $n$ are superscripts that indicate, respectively, the charged and neutral nucleon-Roper reactions. Our conventions are that $F_{1(2),u}^{\ast}$ and $F_{1(2),d}^{\ast}$ refer to the $u$- and $d$-quark contributions to the equivalent Dirac (Pauli) form factors of the $\gamma p\to R^+$ reaction, and the results are normalised such that the \emph{elastic} Dirac form factors of the proton and charged-Roper yield $F_{1u}(Q^2=0)=2$, $F_{1d}(Q^2=0)=1$, thereby ensuring that these functions count $u$- and $d$-quark content in the bound-states.

\begin{figure}[t]
\centerline{%
\includegraphics[clip,width=0.40\textwidth]{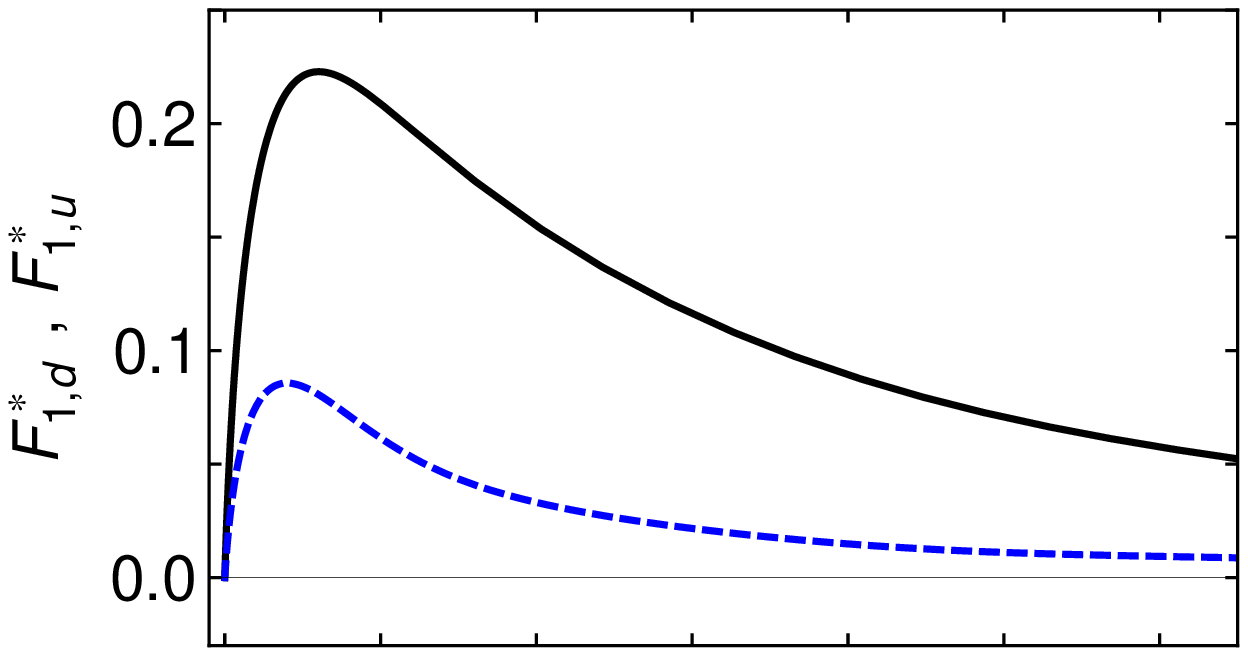}}
\vspace*{-11.3ex}

\centerline{%
\includegraphics[clip,width=0.418\textwidth]{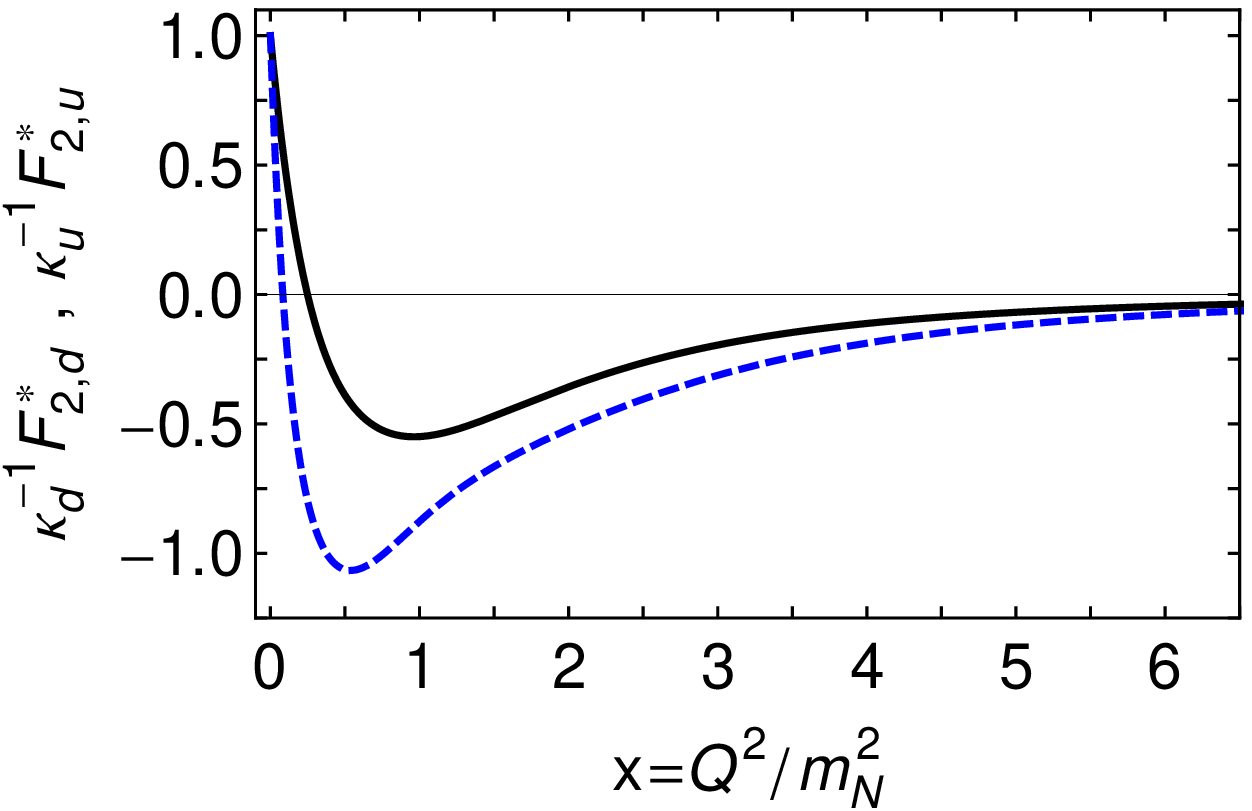}\hspace*{2.2ex}}
\caption{\label{figFud}
Flavour separation, $\gamma\,p\to R^+$ transition form factors: $u$-quark, solid black; and $d$-quark, dashed blue.  \emph{Upper panel} -- Dirac transition form factor.  \emph{Lower panel} -- Pauli transition form factor, with $\kappa_u^\ast  = F_{2,u}^\ast(0) = -0.91$, $\kappa_d^\ast = F_{2,d}^\ast(0) = 0.14$.
}
\end{figure}

\begin{figure}[t]
\centerline{%
\includegraphics[clip,width=0.40\textwidth]{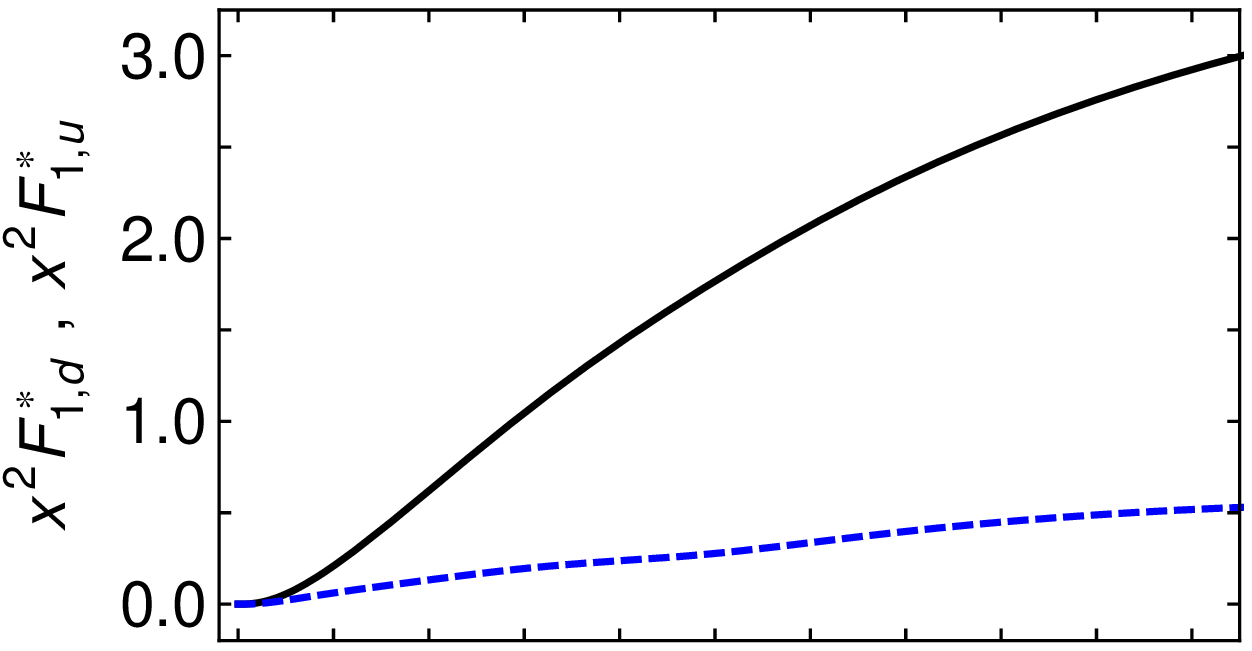}}
\vspace*{-11.1ex}

\centerline{%
\includegraphics[clip,width=0.416\textwidth]{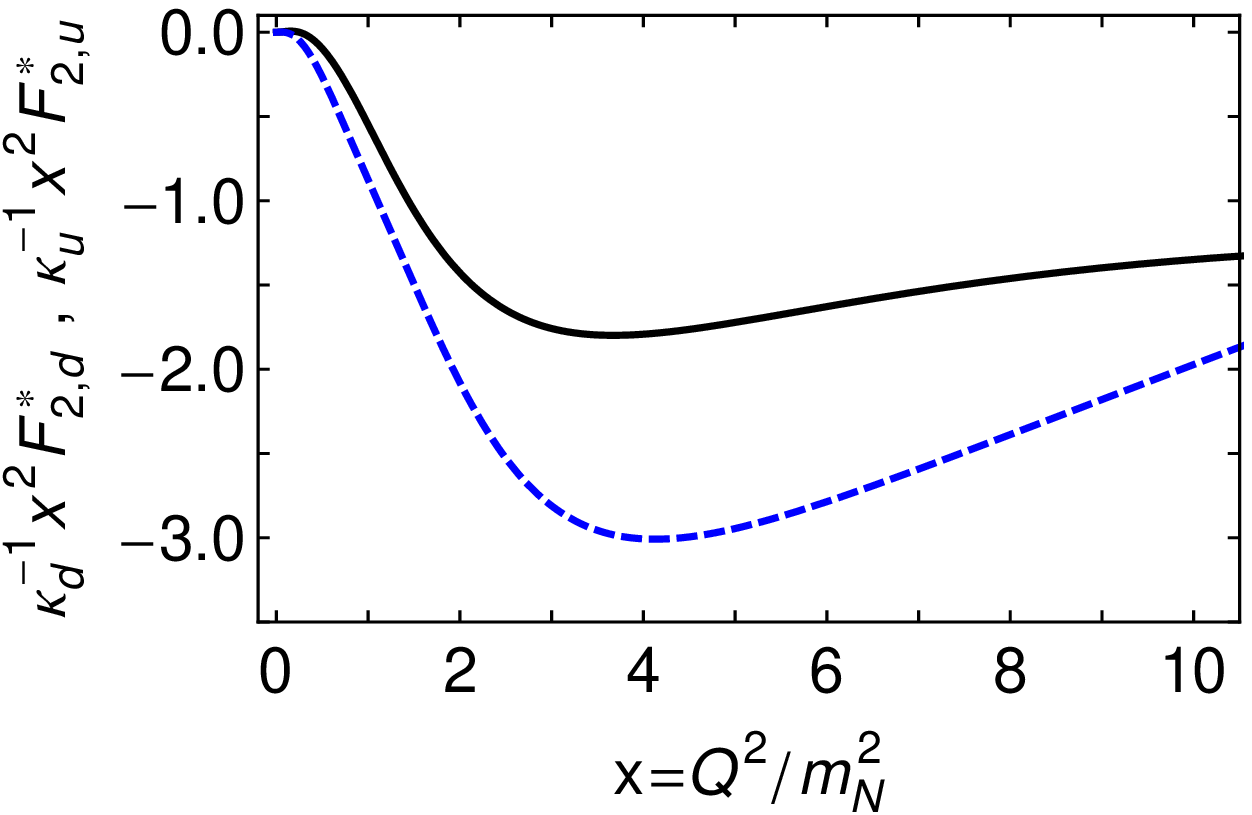}\hspace*{2.2ex}}
\caption{\label{figxFud}
$x^2=(Q^2/m_N^2)^2$-weighted behaviour of the flavour separated $\gamma\,p\to R^+$ transition form factors: $u$-quark, solid black; and $d$-quark, dashed blue.  \emph{Upper (lower) panel} -- Dirac (Pauli) transition form factor.
}
\end{figure}

We computed the $\gamma\,n\to R^0$ transition form factors, using the framework in Ref.\,\cite{Segovia:2015hra}, and employed Eqs.\,\eqref{FlavourSep} to determine the flavour separation of the nucleon-Roper transition.  The results are depicted in Figs.\,\ref{figFud}, \ref{figxFud}.

The upper panels of Figs.\,\ref{figFud}, \ref{figxFud}, depicting the flavour-separated Dirac transition form factor, show an obvious similarity to the analogous form factor determined in elastic scattering: the $d$-quark contribution is less-than half the $u$-quark contribution for momenta sufficiently far outside the neighbourhood of $x_N=0$ within which they both vanish; and the $d$-quark contribution falls more rapidly after their almost coincident maxima.  The noticeable difference, however, is the absence of a zero in $F_{1,d}^{\ast}$, which is a salient feature of the analogous proton elastic form factor.

The lower panels of Figs.\,\ref{figFud}, \ref{figxFud} depict the  flavour-separated Pauli transition form factor.  In this instance the similarities are less obvious, but they are revealed once one recognises that the rescaling factors satisfy $|\kappa_d^\ast/\kappa_u^\ast | < \tfrac{1}{6}$ \emph{cf}.\ a value of $\sim \tfrac{2}{5}$ in the elastic case \cite{Cloet:2008re}.  Accounting for this, the behaviour of the $u$- and $d$-quark contributions to the charged-Roper Pauli transition form factor are comparable with the kindred contributions to the elastic form factor, especially insofar as the $d$-quark contribution falls dramatically on $x\gtrsim 4$ whereas the $u$-quark contribution evolves more slowly.

An explanation for the pattern of behaviour in Figs.\,\ref{figFud}, \ref{figxFud} is much the same as that for the analogous proton elastic form factors \cite{Segovia:2015ufa} because the diquark content of the proton and its first radial excitation are almost identical.  In both systems, the dominant piece of the associated Faddeev wave functions is $\psi_0$, namely a $u$-quark in tandem with a $[ud]$ (scalar diquark) correlation, which produces 62\% of each bound-state's normalisation \cite{Segovia:2015hra}.  If $\psi_0$ were the sole component in both the proton and charged-Roper, then $\gamma$--$d$-quark interactions would receive a $1/x_N$ suppression on $x_N>1$, because the $d$-quark is sequestered in a soft correlation, whereas a spectator $u$-quark is always available to participate in a hard interaction.  At large $x_N$, therefore, scalar diquark dominance leads one to expect $F^\ast_d \sim F^\ast_u/x_N$.  Naturally, precise details of this $x_N$-dependence are influenced by the presence of pseudovector diquark correlations in the initial and final states, which guarantees that the singly-represented quark, too, can participate in a hard scattering event, but to a lesser extent.

The infrared behaviour of the flavour-separated $\gamma p \to R^+$ transition form factors owes to a complicated interference between the influences of orthogonality, which forces $F^\ast_{1,u}(x_N=0)=0=F^\ast_{1,d}(0)$, and quark-core and MB\,FSI contributions.  However, whilst the latter pair act in similar ways for both elastic and transition form factors, orthogonality is a fundamental difference between the two processes and it is therefore likely to be the dominant effect at infrared momenta.

The information contained in Figs.\,\ref{figF1} -- \ref{figxFud} provides clear evidence in support of the notion that many features in the measured behaviour of $\gamma N \to R$ electromagnetic transition form factors are primarily driven by the presence of strong diquark correlations in the nucleon and its first radial excitation.  In our view, inclusion of a ``meson cloud'' cannot qualitatively affect the salient features of these transition form factors, any more than it does the analogous nucleon elastic form factors \cite{Cloet:2012cy, Cloet:2014rja}.


%
\smallskip

\noindent\textbf{5.$\;$Epilogue}.
Dynamical chiral symmetry breaking (DCSB) entails the presence of strong, nonpointlike, interacting diquark correlations in the nucleon and its resonances.  Their existence has numerous observable consequences, an array of which are detailed above in connection with transitions of the nucleon to its two lowest-lying excitations.  
These signals are most prominent for momentum transfers $Q^2 \gtrsim 2\,$GeV$^2$, whereupon contributions from meson-baryon final state interactions are typically negligible.
Precise measurements already exist \cite{Aznauryan:2009mx, Aznauryan:2012ec, Aznauryan:2012ba},
novel experiments are approved at JLab\,12, and others are either planned or under consideration as part of an international effort to measure transition electrocouplings of all prominent nucleon resonances \cite{Aznauryan:2012ba, E12-09-003, E12-06-108A}.  Our predictions can therefore be thoroughly tested in the foreseeable future, and such efforts have the potential to deliver empirical information that would address a wide range of issues, including, \emph{e.g}.: is there an environment sensitivity of DCSB; and are quark-quark correlations an essential element in the structure of all baryons?

\begin{acknowledgments}
We thank R.~Gothe, V.~Mokeev and V.~Burkert for suggesting this problem, and T.S.-H.\,Lee and T.\,Sato for numerous informative discussions.
This work was supported by:
the Alexander von Humboldt Foundation;
and the U.S. Department of Energy, Office of Science, Office of Nuclear Physics, under contract no. DE-AC02-06CH11357.
\end{acknowledgments}



\end{document}